\documentclass[aps,prl,reprint,showpacs,superscriptaddress]{revtex4-1}
\usepackage{amsmath}
\usepackage{amsfonts}
\usepackage{amssymb}
\usepackage{color}
\usepackage{blindtext}
\usepackage{framed}
\usepackage{graphicx}
\usepackage{bm}
\usepackage{natbib}
\usepackage{latexsym}
\usepackage[final]{changes}
\usepackage{braket}
\usepackage{url}

\begin{document}
\title{Role of the Plasmoid Instability in Magnetohydrodynamic Turbulence}

\author{Chuanfei Dong}
\email{dcfy@princeton.edu}
\affiliation{Department of Astrophysical Sciences, Princeton University, Princeton, NJ 08544, USA}
\affiliation{Princeton Plasma Physics Laboratory, Princeton University, Princeton, NJ 08540, USA}
\author{Liang Wang}
\affiliation{Department of Astrophysical Sciences, Princeton University, Princeton, NJ 08544, USA}
\affiliation{Princeton Plasma Physics Laboratory, Princeton University, Princeton, NJ 08540, USA}
\author{Yi-Min Huang}
一\affiliation{Princeton Plasma Physics Laboratory, Princeton University, Princeton, NJ 08540, USA}
\author{Luca Comisso}
\affiliation{Department of Astrophysical Sciences, Princeton University, Princeton, NJ 08544, USA}
\affiliation{Princeton Plasma Physics Laboratory, Princeton University, Princeton, NJ 08540, USA}
\affiliation{Department of Astronomy, Columbia University, New York, NY 10027, USA}
\affiliation{Columbia Astrophysics Laboratory, Columbia University, New York, NY 10027, USA}
\author{Amitava Bhattacharjee}
\affiliation{Department of Astrophysical Sciences, Princeton University, Princeton, NJ 08544, USA}
\affiliation{Princeton Plasma Physics Laboratory, Princeton University, Princeton, NJ 08540, USA}


\begin{abstract}
The plasmoid instability in evolving current sheets has been widely studied due to its effects on the disruption of current sheets, the formation of plasmoids, and the resultant fast magnetic reconnection. In this Letter, we study the role of the plasmoid instability in two-dimensional magnetohydrodynamic (MHD) turbulence by means of high-resolution direct numerical simulations. At sufficiently large magnetic Reynolds number ($R_m=10^6$), the combined effects of dynamic alignment and turbulent intermittency lead to a copious formation of plasmoids in a multitude of intense current sheets. The disruption of current sheet structures facilitates the energy cascade towards small scales, leading to the breaking and steepening of the energy spectrum. In the plasmoid-mediated regime, the energy spectrum displays a scaling that is close to the spectral index $-2.2$ as proposed by recent analytic theories. We also demonstrate that the scale-dependent dynamic alignment exists in 2D MHD turbulence and the corresponding slope of the alignment angle is close to 0.25. 
\end{abstract}


\maketitle

{\it Introduction.} Magnetohydrodynamic (MHD) turbulence plays a fundamental role in the transfer of energy in a wide range of space and astrophysical systems, from the solar corona \cite{Matthaeus99,Cranmer2007} and accretion disks \cite{BalbusHawley98,BrandSubra05}, to the interstellar medium \cite{Armstrong95,Elmegreen04} and galaxy clusters \cite{Zweibel97,Miniati15}. Indeed, in all these environments, MHD turbulence is responsible for the transfer of energy from the large scales where energy is provided to the small scales where it is dissipated. Establishing a detailed understanding of MHD turbulence is a pivotal but yet unresolved problem, with far-reaching repercussions in many research areas.

An important feature of MHD turbulence is the tendency to develop sheets of strong electric current density \cite{Biskamp89,Politano95,Biskamp2003,Biskamp2001,Mininni06,Zhdankin13,Wan14,Makwana15,Matthaeus15,Mak17,Yang17}. 
These current sheets are natural sites of magnetic reconnection, leading to the formation of plasmoids that eventually disrupt the sheet-like structures in which they are born \cite{ML86,Politano89,Goldstein95,Servidio09,Servidio11,Wan2013,HuBha16,Cerri2017,Franci17,Franci18}. The first analytic calculation of the impact of plasmoid formation on the MHD turbulent cascade is attributable to Carbone, Veltri and Mangeney \cite{Carbone90}, who proposed that current sheet structures in a turbulent environment disrupt when $\gamma \tau_{\rm{nl}} \sim 1$, with $\tau_{\rm{nl}}$ and $\gamma$ corresponding to the nonlinear eddy turnover time and the growth rate of the fastest tearing mode, respectively. Under this condition, they derived a length scale at which the inertial range of turbulence breaks and found that the energy spectrum steepens because of the plasmoid instability. Interestingly, by performing hybrid-kinetic simulations, Refs. \citep{Cerri2017,Franci17} showed that plasmoids can also directly contribute to the formation of a fully-developed turbulent spectrum across the so-called ion break. \added{To date, there have been several attempts to further predict the disruption criteria and the energy spectrum in the plasmoid-mediated regime \citep{Pucci2014,Comisso2016,Tenerani2016,Comisso2017,Huang2017,Mallet2017,LouBold2017,BoldLou2017,Comisso2018}.}


Despite a series of analytic studies predicting a break of the energy spectrum caused by the plasmoid instability, no definitive evidence has been provided by direct numerical simulations of MHD turbulence so far. This is mainly due to the limitation of computational resources enabling the high magnetic Reynolds numbers that are required to achieve a regime where the plasmoid formation is statistically significant to affect the turbulent energy cascade. This Letter aims to address this problem by performing numerical simulations at unprecedented large magnetic Reynolds numbers (up to $R_m = 10^6$). Although recent theoretical arguments are based on three-dimensional turbulence, we will consider a two-dimensional scenario. In this way, we can achieve high resolution to resolve the plasmoid instability associated with the high $R_m$ and observe a concomitant alteration in the spectrum of the turbulent cascade. According to our simulations, we demonstrate that plasmoids can cause a steepening of the energy spectrum, which exhibits a slope that is close to a value of $-2.2$. 

{\it Method.} The governing equations of our numerical model are the dimensionless visco-resistive MHD equations:
\begin{equation} \label{continuity_eq}
\partial_t \rho  + \nabla \cdot (\rho {\bf{u}}) = 0 \, ,
\end{equation}
\begin{equation} \label{}
\partial_t (\rho {\bf{u}}) + \nabla  \cdot (\rho {\bf{uu}}) =  - \nabla \left( p + B^2/2 \right) + \nabla  \cdot ({\bf{BB}}) + \nu {\nabla ^2}(\rho {\bf{u}}) \, ,
\end{equation}
\begin{equation} \label{}
\partial_t p + \nabla  \cdot (p {\bf{u}}) = (\gamma - 1) \left( { - p \nabla \cdot {\bf{u}} + \eta {\bf{J}}^2} \right) \, ,
\end{equation}
\begin{equation} \label{induction_eq}
\partial_t {\bf{B}} = \nabla  \times ({\bf{u}} \times {\bf{B}} - \eta {\bf{J}}) \, ,
\end{equation}
where $\rho$, ${\bf{u}}$ and $p$ are the mass density, velocity, and pressure of the plasma, respectively; ${\bf{B}}$ is the magnetic field and ${\bf{J}} = \nabla  \times {\bf{B}}$ denotes the electric current density. The kinematic viscosity and the magnetic diffusivity are denoted  as $\nu$ and $\eta$, respectively, while $\gamma$ is the adiabatic index.

We solve Eqs.~(\ref{continuity_eq})-(\ref{induction_eq}) using the BATS-R-US MHD code \cite{Toth2012} in a domain $\{(x,y):-L_0/2 \leq x, y \leq L_0/2\}$, where $L_0$ is set to unity. Periodic boundary conditions are employed in both directions. Lengths are normalized to the box size $L_0$, velocities to the characteristic Alfv\'{e}n speed $V_A$, and time to $L_0/V_A$. We initialize the simulations by placing uncorrelated, equipartitioned velocity and magnetic field fluctuations in Fourier harmonics. More specifically, we set the initial velocity $u_x=\sum a_{mn}n\sin(k_m x+\xi_{mn})\cos(k_n y+\zeta_{mn})$ and $u_y=\sum -a_{mn}m\cos(k_m x+\xi_{mn})\sin(k_n y+\zeta_{mn})$, where $m$ and $n$ indicate the mode numbers in $x$ and $y$ directions; the wavenumber $k_m=2\pi m/L_0$, and $\xi_{mn}$ and $\zeta_{mn}$ are random phases. Energy is initialized in the range $l_{\rm{min}} \leq (m^2+n^2)^{1/2} \leq l_{\rm{max}}$, where $l_{\rm{min}}=1$ and $l_{\rm{max}}=10$ with $a_{mn}=u_0/[(m^2+n^2)(l_{\rm{max}}^2-l_{\rm{min}}^2)]^{1/2}$. We set the magnetic field in the same way, with the constant $u_0$ replaced by $B_0$ and with different random phases. The out-of-plane magnetic field is set to zero; hence, there is no mean magnetic field. The constants $u_0$ and $B_0$ determine the strength of the initial velocity and magnetic fields. We set $u_0=B_0=2$, which gives the initial energy $E = \frac{1}{2}\langle |\bm{u}|^2 +  |\bm{B}|^2 \rangle \simeq \frac{1}{8}(u_0^2+B_0^2) = 1$, where $\langle ... \rangle$ represents the spatial average. The plasma density and pressure are initially set to  constant values $\rho=1$ and $p = 10$, respectively.

We performed simulations with different values ($4\times10^4$, $8\times10^4$, $1\times10^5$, $2\times10^5$, $1\times10^6$) of magnetic Reynolds number, given by $R_m = u_0 L_0 /\eta$. We vary the numbers of grid points from $2000^2$ to $64000^2$ for convergence study, ensuring the resolution is high enough to resolve the plasmoid instability associated with the highest $R_m$. Our analyses are based upon two cases: $R_m = 8 \times 10^4$ and $R_m = 1 \times 10^6$, with $64000^2$ grid points. For the two cases, the magnetic diffusivity and viscosity are chosen to be (1) $\eta=2\times10^{-6}$, $\nu=1\times10^{-6}$ and (2) $\eta=2.5\times10^{-5}$, $\nu=1.25\times10^{-5}$; therefore, the magnetic Prandtl number $P_m=\nu/\eta=0.5$. We analyze our data from the snapshot near the peak of the mean-square current density $\langle J_z^2 \rangle$, when the turbulence is fully developed ($t \sim 0.2$).

\begin{figure}
\begin{center}
\includegraphics[width=8.8cm]{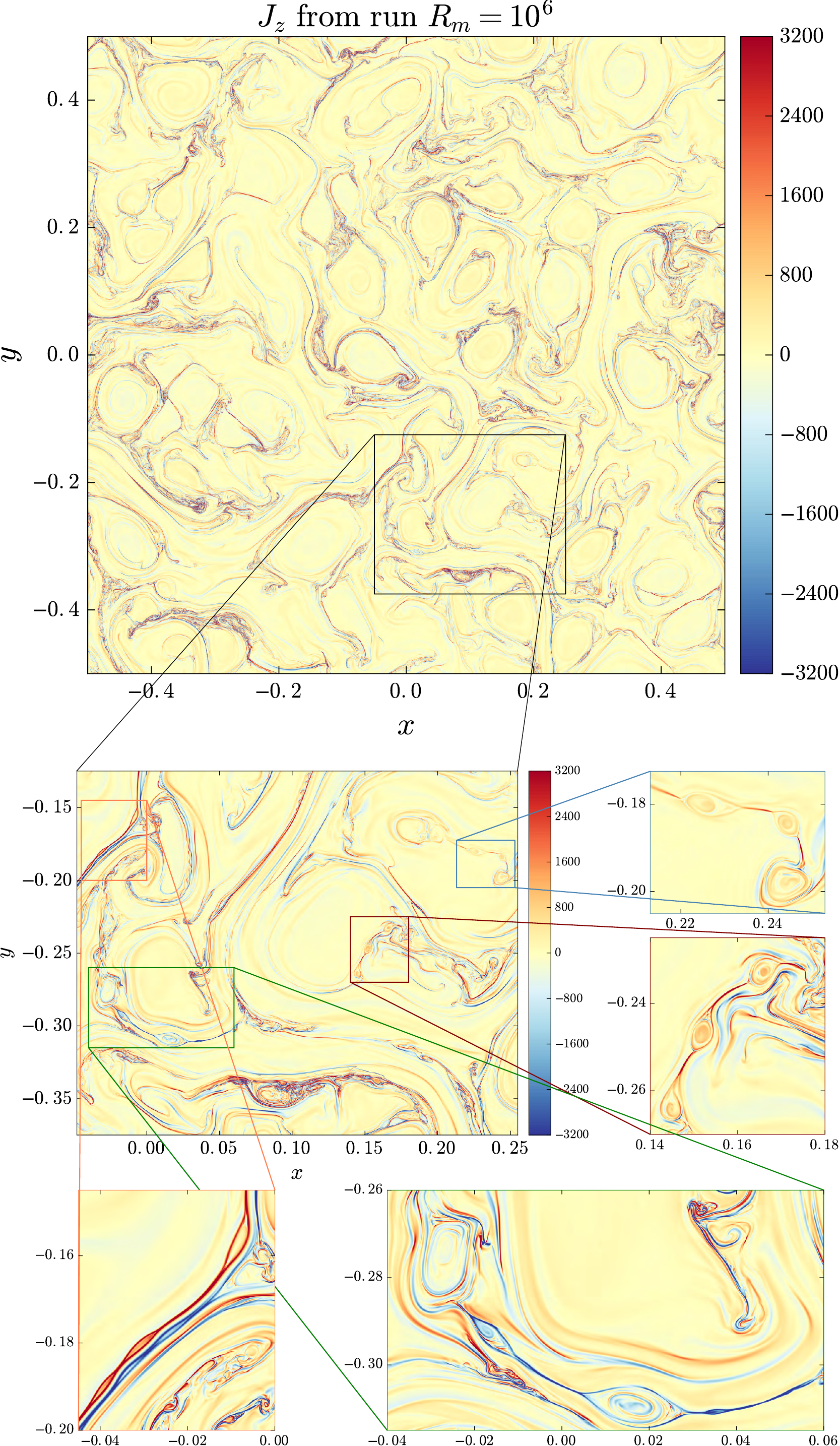}
\end{center}
\caption{2D contour plots of the current density $J_z$ at $t=0.2$ for the simulation with magnetic Reynolds number $R_m=1\times10^6$. Zoomed-in subdomains are used for the illustration of plasmoids. Copious formation of plasmoids occurs in multiple intense current sheets because of the plasmoid instability.}
\label{fig1}
\end{figure}

\begin{figure}
\begin{center}
\includegraphics[width=8.8cm]{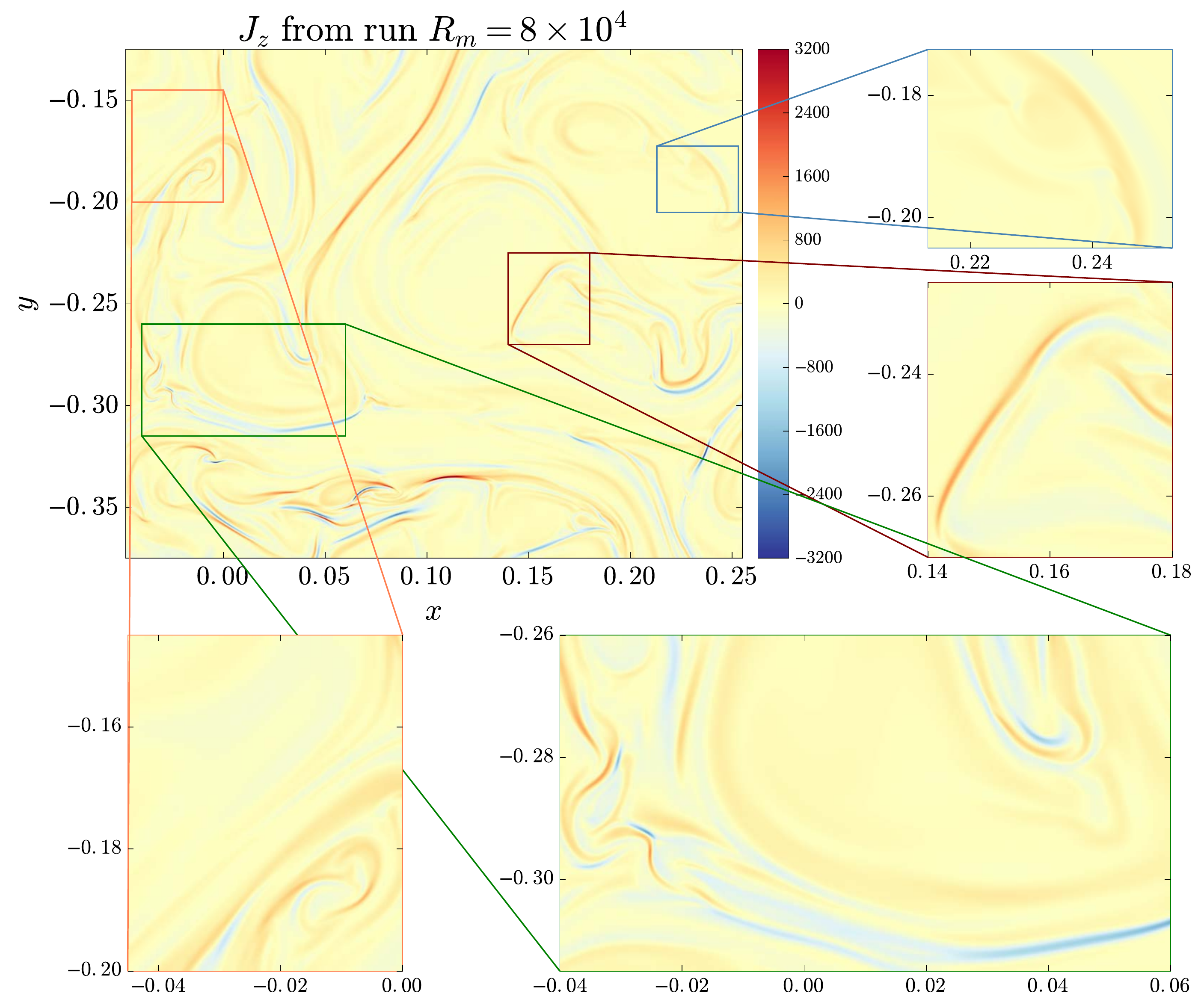}
\end{center}
\caption{2D contour plots of the current density $J_z$ (by choosing the same zoomed-in boxes and identical colorbar range as that of Fig.~\ref{fig2}) at $t=0.2$ for the simulation with magnetic Reynolds number $R_m=8\times10^4$. Barely any plasmoids can be observed in the current sheets due to the relatively low $R_m$ that prevents the current sheet from further thinning, therefore, less intense current sheets develop in this simulation.}
\label{fig2}
\end{figure}

{\it Results.} Fig.~\ref{fig1} depicts the simulated $J_z$ with $R_m=1 \times 10^6$, and displays coherent structures with intermittent intensity. In order to show plasmoids formed within the time-evolving current sheets \citep[e.g.,][]{Comisso2016,Huang2017,Comisso2017}, we zoom into the selected regions. \added{It is noteworthy that the plasmoids grow locally within the current sheets in our simulation \cite{animation}, instead of being convected into the current sheets by finite amplitude fluctuations \cite{ML86,Wan2013}.} Chains of plasmoids are formed at different evolution stages in multiple intense current sheets. As will be discussed below, the disruption of the current sheet structures due to plasmoids can alter the turbulence energy spectrum by facilitating the energy cascade toward small scales. In contrast, Fig.~\ref{fig2}, where $R_m=8\times10^4$, reveals that plasmoids barely form in the same zoomed-in regions. The current sheet intensity for the low $R_m$ case is significantly lower than that of the high $R_m$ case. This is because the low $R_m$ (equivalent to the Lundquist number, $S= V_A L_0 /\eta$, in this study) prevents the current sheet from further thinning to smaller scales.

\begin{figure}
\begin{center}
\includegraphics[width=8.8cm]{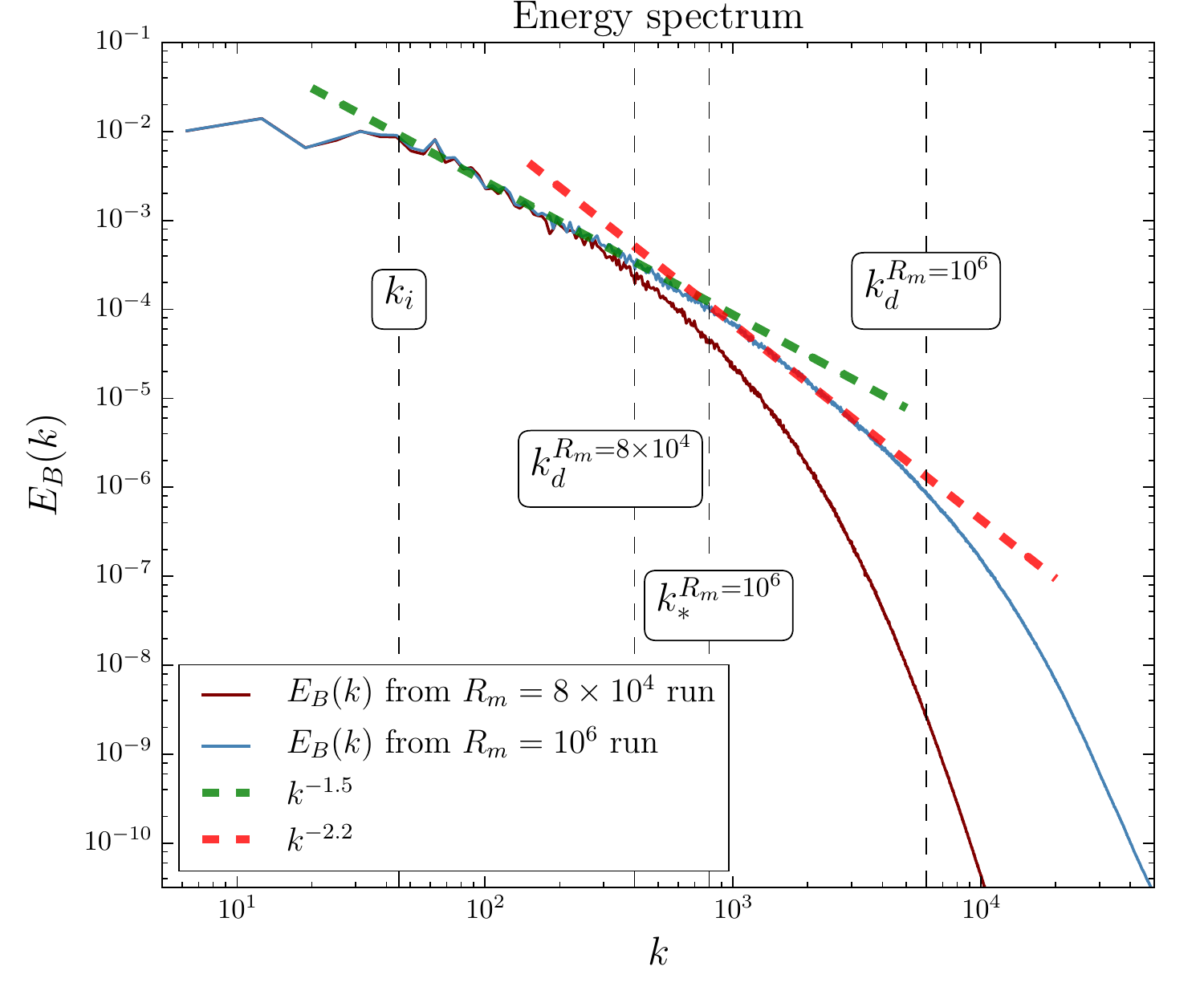}
\end{center}
\caption{Magnetic energy spectrum of the MHD turbulent cascade. $k_i$ is approximately the start of the inertial range, $k_d$ represents the dissipation range, and $k_*$ indicates where the spectrum breaks and thus the start of the sub-inertial range. Both $R_m$ cases share the same $k_i$. In the low $R_m$ case, $k_*$ is absent or indistinguishable from $k_d$ due to the short scale separation. We choose $k_i$ based on the left edge of the fitted power-law spectrum. $k_*$ is determined by the intersection of the two fitted power-law spectra \cite{Note1}. We estimate $k_d$ by the condition that $\eta\int_{0}^{k_d} E_B(k)k^2 dk$ accounts for approximately half of the resistive dissipation power $\eta \braket{J^2}$.}
\label{fig3}
\end{figure}

We now examine the turbulence energy cascade by computing the magnetic energy spectra, as shown in Fig. \ref{fig3}. The start of the inertial range, the transition scale between the standard inertial range and the plasmoid-mediated range, and the dissipation scale are denoted as $k_i$, $k_*$, and $k_d$, respectively. The low and high $R_m$ simulations share the same $k_i$ as a result of the identical initial setup except the distinct $\eta$ and $\nu$. In the low $R_m$ case, $k_*$ is absent or indistinguishable from $k_d$ due to the small scale separation. Compared with the high $R_m$ simulation, the spectrum of the low $R_m$ case has a shorter inertial range and falls more quickly into the dissipation range. For the longer inertial range of the high $R_m$ case, the formation of the plasmoids breaks the energy spectrum. The standard inertial range is characterized by a spectral index of $-1.5$, in accordance with \cite{Iroshnikov63,Kraichnan65}. For $k>k_*$ the spectrum becomes steeper and close to a spectral index of $-2.2$ \cite{Note2}, as recently proposed in Refs. \cite{BoldLou2017,Comisso2018} (also see Ref. \cite{Mallet2017}, where the spectrum was proposed to be bounded between the spectral indexes -5/3 and -2.3).
This sub-inertial range is only displayed in the high $R_m$ case due to the copious formation of plasmoids that characterizes this simulation. The co-occurrence of both plasmoids and steeper spectral index in the high $R_m$ case is a reasonable indication that the steepening of the energy spectrum results from the disruption of the current sheet structures caused by plasmoids, which facilitates the energy cascade toward small scales.

\begin{figure}
\begin{center}
\includegraphics[width=8.8cm]{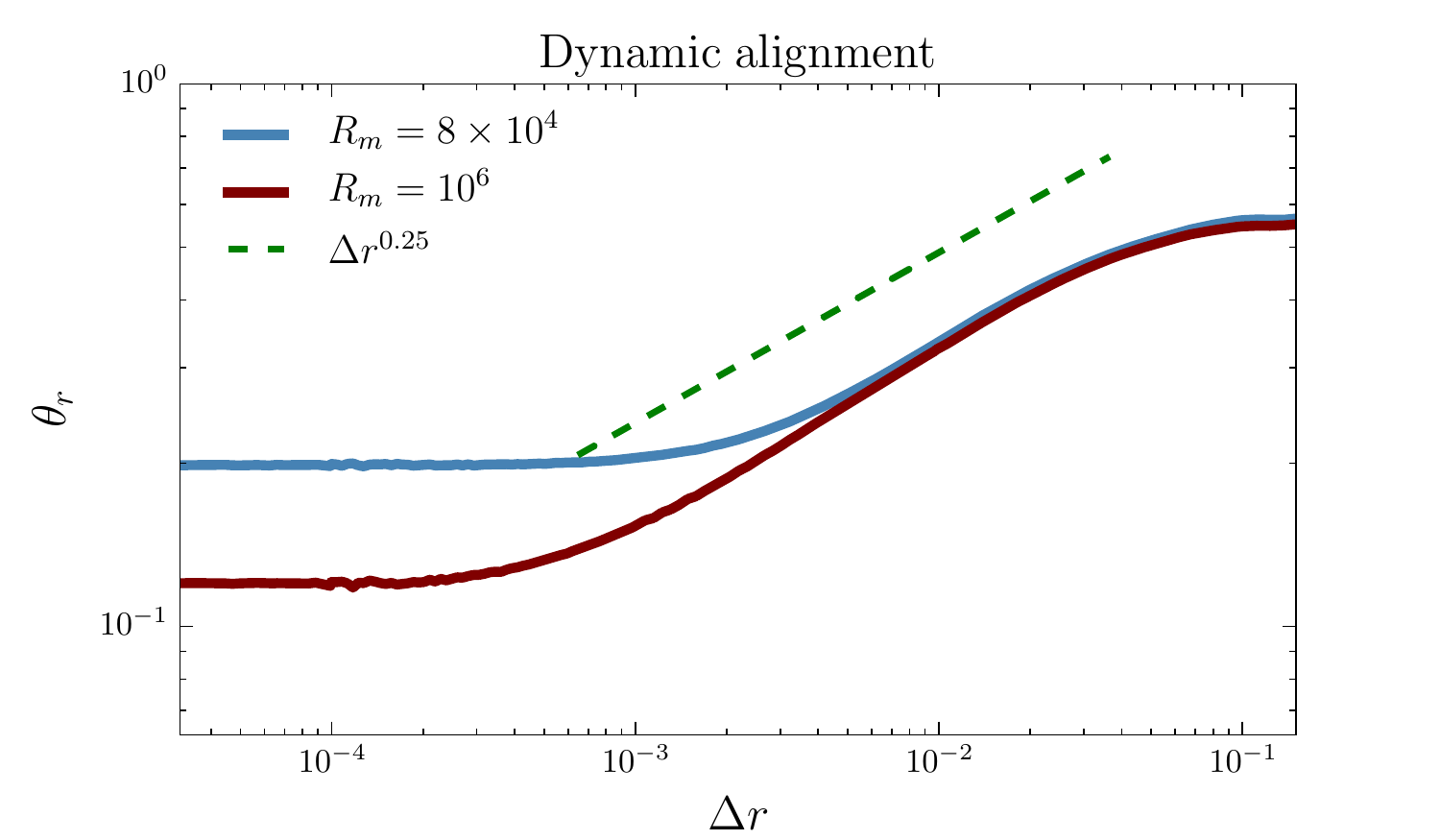}
\end{center}
\caption{The alignment angle $\theta_r$ in a log-log scale as a function of the spatial separation $\Delta r$. The dashed line indicates the scaling law $\theta_r \propto \Delta r^{0.25}$. The higher $R_m$ case has a smaller alignment angle $\theta_r$, especially at small spatial separations.}
\label{fig4}
\end{figure}

Previously, the tendency of forming elongated, plasmoid-prone current sheets at small scales has been attributed to the effect of dynamic alignment \cite{Boldyrev2006}, with an underlying assumption that the alignment angle can be employed as a proxy for current sheet inverse aspect ratio \cite{Mallet2017,LouBold2017,Comisso2018}. We measure the dynamic alignment angle $\theta_r$ as a function of the spatial separation between two sampling points, $\Delta r$, according to the definition $\sin \theta_r = \left\langle |{\delta {\bf{u}} \times \delta {\bf{B}}}| \right\rangle /\left\langle {|\delta {\bf{u}}||\delta {\bf{B}}|} \right\rangle $ given by Ref.~\citep{Mason2006}. Here,  $\delta \bf{B}=\bf{B}(\bf{r}+\Delta \bf{r})-\bf{B}(r)$ is the difference in the magnetic field between two points randomly sampled throughout the entire domain; likewise, $\delta \bf{u}$ is the difference in the velocity. The results, shown in Fig.~\ref{fig4}, indicate that the alignment angle is close to $\theta_r \propto \Delta r^{0.25}$ for a wide range of $\Delta r$. This scaling relation is analogous to the one that characterizes 3D MHD turbulence \cite{Boldyrev2006}, implying that the dynamic alignment may be universal in MHD turbulence regardless of the spatial dimensions. At a small spatial separation $\Delta r$, the higher $R_m$ simulation is distinguished by a smaller $\theta_r$ than the low $R_m$ one. This indicates that the high $R_m$ case is associated with a higher current sheet aspect ratio based on statistics as expected. 

\begin{figure}
\begin{center}
\includegraphics[width=8.6cm]{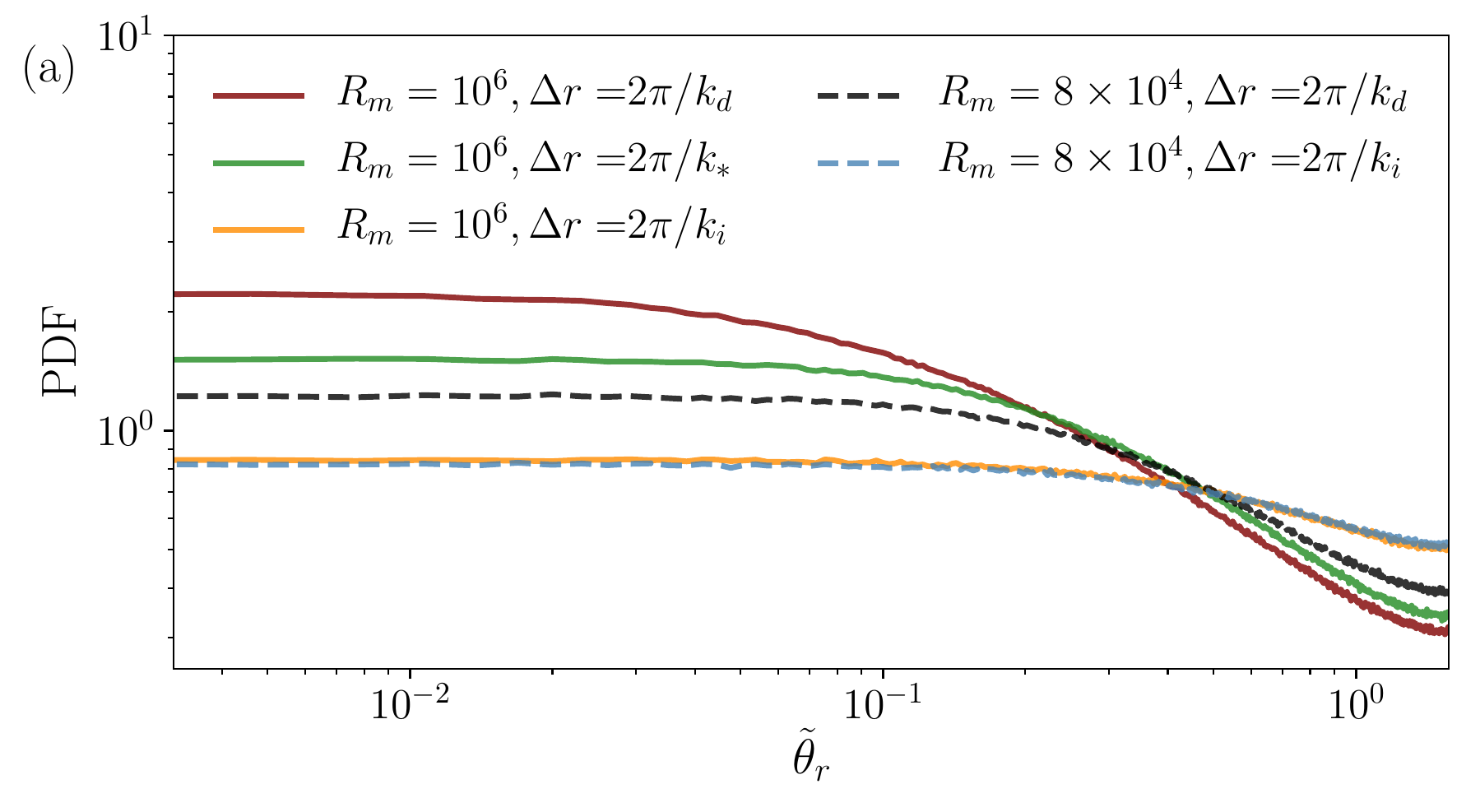}
\includegraphics[width=8.8cm]{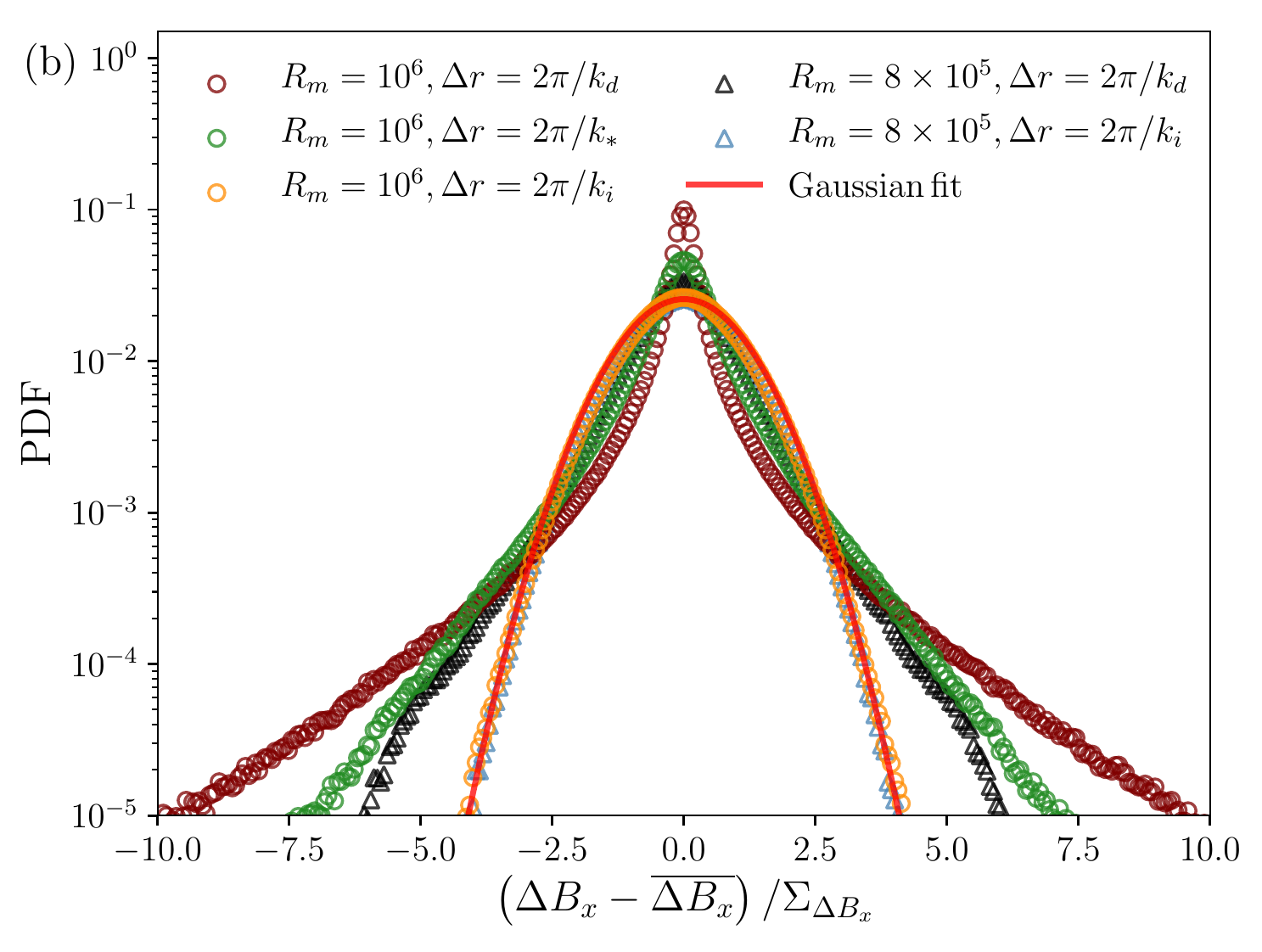}
\end{center}
\caption{PDFs of (a) the alignment angle $\widetilde{\theta_r}$ and (b) $(\Delta B_x- \overline{\Delta B_x})/\Sigma_{\Delta B_x}$ ($\Sigma_{\Delta B_x}$ the standard deviation of $\Delta B_x$) at different separations determined by $k_i$, $k_*$ and $k_d$ as indicated in Fig.~\ref{fig3}. (a) $\widetilde{\theta_r}$ has a relatively uniform distribution at the $k_i$ scale for both $R_m$ cases. At the $k_d$ scale, the high $R_m$ case exhibits higher nonuniformity than the low $R_m$ case. (b) At the $k_i$ scale, both $R_m$ cases show similar PDFs that can accurately fit the Gaussian distribution. The PDFs at $k_d$ between low and high $R_m$ cases are clearly distinguishable; the latter deviates more from the Gaussian distribution.}
\label{fig5}
\end{figure}

Although the alignment angle progressively diminishes as the spatial separation decreases, the rather moderate alignment with $\theta_r > 0.1$ yields an aspect ratio no more than $O(10)$, which appears too low to cause the onset of the plasmoid instability. On the other hand, the existence of elongated and intense current sheets (see Fig.~\ref{fig1}), despite the moderate values of the dynamic alignment, indicates that intermittency may play a key role \cite{Biskamp2003}. In particular, MHD turbulence is known to be more intermittent in 2D than in 3D \cite{Biskamp2001}. Given that Fig.~\ref{fig4} only presents an averaged picture of the dynamic alignment, we further define an alignment angle $\widetilde{\theta_r}$ for an individual pair of $\delta {\bf{u}}$ and $\delta {\bf{B}}$  with $\sin \widetilde{\theta_r} = |\delta {\bf{u}} \times \delta {\bf{B}}| /(|\delta {\bf{u}}||\delta {\bf{B}}|)$. Fig.~\ref{fig5} (a) depicts the probability density functions (PDFs) of $\widetilde{\theta_r}$ calculated at separations determined by $k_i$, $k_*$ and $k_d$ as indicated in Fig.~\ref{fig3}. At the scale of $k_i$, the PDFs of both $R_m$ cases have a relatively uniform distribution spanning over all the alignment angles, indicating the alignment angle at the $k_i$ scale is nearly isotropic. As the separation $\Delta r$ decreases, the nonuniformity in the PDF gets amplified and the mean value of the alignment angle shifts to smaller $\widetilde{\theta_r}$. At the $k_d$ scale, the higher $R_m$ case exhibits more anisotropy than the lower $R_m$ case. Notably,  there exists a substantial probability for the individual alignment angle $\tilde{\theta_r}$ to be significantly smaller than the averaged alignment angle $\theta_r$ shown in Fig.~\ref{fig4}.

To further quantify intermittency, we analyze the PDFs of magnetic field increments $|\Delta_{\Delta r}\mathbf{B}| = |\mathbf{B}(\bf{r}+\Delta \bf{r}) - \mathbf{B}(\bf{r})|$ for various $\Delta r$ \cite{Greco2008}. In Fig.~\ref{fig5} (b), we show the PDFs of the $x$ component $\Delta_{\Delta r}\mathbf{B_x}$ at the same separations as Fig.~\ref{fig5} (a). PDFs of $\Delta_{\Delta r}\mathbf{B_y}$, not shown here, are analogous to that of $\Delta_{\Delta r}\mathbf{B_x}$. The PDFs at the dissipation scale, $k_d$, between the two $R_m$ cases are highly distinguishable; the high $R_m$ case has a more distinct non-Gaussian distribution. The large deviation from a Gaussian distribution at $k_*$ and $k_d$ scales for the high $R_m$ case indicates that turbulence is highly intermittent at those scales. Both Figs.~\ref{fig5} (a) and ~\ref{fig5}(b) help explain the plasmoid formation in the elongated and intense current sheets despite the statically moderate values of the alignment angle.

{\it Conclusions.} In this Letter, we show, via direct numerical simulations, that the plasmoid instability can modify the MHD turbulent cascade for sufficiently large magnetic Reynolds numbers. For the higher magnetic Reynolds number considered ($R_m=1 \times 10^6$), the energy spectrum steepens relative to the inertial range when the plasmoid instability becomes effective in disrupting current sheets. This occurs at relatively small scales, where the combined effects of dynamic alignment and intermittency produce current sheets more prone to the plasmoid instability. In this plasmoid-mediated regime, we found that the magnetic energy spectrum exhibits a spectral index close to $-2.2$. Therefore, the disruption of current sheets due to the plasmoid instability facilitates the energy cascade toward small scales.

To quantify the effect of the dynamic alignment, we measured the alignment angle $\theta_r$, which decreases toward small scales with a slope that is approximately captured by the scaling $\theta_r \propto \Delta r^{0.25}$. This is analogous to what is found in 3D MHD turbulence, thus suggesting a possible common explanation underlying this behavior. Despite the fact that the dynamic alignment can be related to the increase of the aspect ratio of current sheets, for the magnetic Reynolds number investigated here it is not sufficient to explain the plasmoid formation per se. However, intensified nonuniformity in $\theta_r$ and significantly non-Gaussian distribution of magnetic field increments are shown to occur at smaller scales for the case with a high $R_m$, which may suffice to reach the critical current sheet aspect ratio required for the formation of plasmoids within the typical eddy turnover time. 

The alignment angle $\theta_r$ is not affected by the development of plasmoids in our simulations, in contrast to the theoretical predictions of Refs. \cite{BoldLou2017,Comisso2018}. A possible explanation is that the plasmoids start to form and evolve at moderate values of $\theta_r$ with the assistance of intermittency. The impact of the plasmoid instability in modifying the dynamic alignment could become manifest at larger $R_m$, where the anisotropy becomes significantly larger. Furthermore, it is shown in Ref. \cite{Comisso2018} that the plasmoid-mediated regime cannot be characterized by a true power-law due to the nature of the plasmoid instability in time-evolving current sheets. This phenomenon seems to be captured by our numerical simulations based on the adopted $R_m$. However, significantly larger $R_m$ values are required to see a more clear and broader trend.

Several questions remain open, e.g., the role of dimensionality (2D vs. 3D) on intermittency and energy transfer between scales, the difference between decaying turbulence and forced turbulence, and how to examine different existing theories against numerical simulations.

\begin{acknowledgments}
We acknowledge fruitful discussions with Manasvi Lingam, Yan Yang, Greg Hammett, Yao Zhou, and Takuya Shibayama. We would like to thank the anonymous referees' helpful comments and constructive suggestions. This work is supported by NSF grants AGS-1338944 and AGS-1460169, DOE grants DE-SC0016470 and DE-SC0006670, and NASA grant NNX13AK31G. Resources supporting this work were provided by the NASA High-End Computing (HEC) Program through the NASA Advanced Supercomputing (NAS) Division at Ames Research Center. We also would like to acknowledge high-performance computing support from Cheyenne (doi:10.5065/D6RX99HX) provided by NCAR's CISL, sponsored by NSF, and from Trillian at UNH supported by the NSF Grant PHY-1229408. 
\end{acknowledgments}


\begin{thebibliography}{}

\bibitem{Matthaeus99} W.H. Matthaeus, G.P. Zank, S. Oughton, D.J. Mullan and P. Dmitruk, Astrophys. J. \textbf{523}, L93 (1999)  
 
\bibitem{Cranmer2007} S.R. Cranmer, A.A. van Ballegooijen and R.J. Edgar, Astrophys. J. \textbf{171}, 520 (2007) 
 
\bibitem{BalbusHawley98} S.A. Balbus and J.F. Hawley, Reviews of Modern Physics \textbf{70}, 1 (1998) 

\bibitem{BrandSubra05} A. Brandenburg and K. Subramanian, Phys. Rep. \textbf{417}, 1 (2005).

\bibitem{Armstrong95} J.W. Armstrong, B.J. Rickett and S.R. Spangler, Astrophys. J. \textbf{443}, 209 (1995) 

\bibitem{Elmegreen04} B.G. Elmegreen and J. Scalo, Annu. Rev. Astron. Astrophys. \textbf{42}, 211 (2004).

\bibitem{Zweibel97} E.G. Zweibel and C. Heiles, Nature \textbf{385}, 131 (1997). 

\bibitem{Miniati15} F. Miniati and A. Beresnyak, Nature \textbf{523}, 59 (2015).

\bibitem{Biskamp89}  D. Biskamp and H. Welter, Phys. Fluids B \textbf{1}, 1964 (1989).

\bibitem{Politano95} H. Politano, A. Pouquet and P.L. Sulem, Phys. Plasmas \textbf{2}, 2931 (1995).

\bibitem{Biskamp2003} D. Biskamp, {\it Magnetohydrodynamic Turbulence} (Cambridge University Press, 2003).

\bibitem{Biskamp2001} D. Biskamp and E. Schwarz, Phys. Plasmas, \textbf{8}(7), 3282-3292 (2001).

\bibitem{Mininni06} P.D. Mininni,A. G. Pouquet, and D.C. Montgomery, Phys. Rev. Lett. \textbf{97}, 244503 (2006).

\bibitem{Zhdankin13} V. Zhdankin, D.A. Uzdensky, J.C. Perez and S. Boldyrev, Astrophys. J. \textbf{771}, 124 (2013).

\bibitem{Wan14} M. Wan, A. F. Rappazzo, W.H. Matthaeus, S. Servidio, and S. Oughton Astrophys. J. \textbf{797}, 63 (2014).

\bibitem{Makwana15} K.D. Makwana, V. Zhdankin, H. Li, W. Daughton, and F. Cattaneo, Phys. Plasmas \textbf{22}, 042902 (2015).

\bibitem{Matthaeus15} W.H. Matthaeus, M. Wan, S. Servidio, A. Greco, K.T. Osman, S. Oughton, P. Dmitruk, Phil. Trans. R. Soc. A \textbf{373}, 20140154 (2015).

\bibitem{Mak17}  J. Mak, S.D. Griffiths, and D.W. Hughes, Phys. Rev. Fluids \textbf{2}, 113701 (2017).

\bibitem{Yang17}  Y. Yang, W.H. Matthaeus, Y. Shi, M. Wan, and S. Chen, Phys. Fluids  \textbf{29}, 035105 (2017).

\bibitem{ML86} W.H. Matthaeus and S.L. Lamkin, Phys. Fluids \textbf{29}, 2513 (1986).

\bibitem{Politano89} H. Politano, A. Pouquet and P.L. Sulem, Phys. Fluids B \textbf{1}, 2330 (1989).

\bibitem{Goldstein95} M. L. Goldstein, D. A. Roberts, and W. H. Matthaeus, Annu. Rev. Astron. Astrophys. \textbf{33}, 283 (1995).

\bibitem{Servidio09} S. Servidio, W.H. Matthaeus, M.A. Shay, P.A. Cassak and P. Dmitruk, Phys. Rev. Lett. \textbf{102}, 115003 (2009).

\bibitem{Servidio11} S. Servidio, P. Dmitruk, A. Greco, M. Wan, S. Donato, P.A. Cassak, M.A. Shay, V. Carbone and W.H. Matthaeus, Nonlin. Processes Geophys. \textbf{18}, 675 (2011).

\bibitem{Wan2013} M. Wan, W.H. Matthaeus, S. Servidio and S. Oughton, Phys. Plasmas \textbf{20}, 042307 (2013).

\bibitem{HuBha16} Y.-M. Huang and A. Bhattacharjee, Astrophys. J. \textbf{818}, 20 (2016).

\bibitem{Cerri2017} S.S. Cerri and F. Califano, New J. Phys. \textbf{19}, 025007 (2017).

\bibitem{Franci17} L. Franci, S.S. Cerri, F. Califano, S. Landi, E. Papini, A. Verdini, L. Matteini, F. Jenko and P. Hellinger, Astrophys. J. Lett. \textbf{850}, L16 (2017).

\bibitem{Franci18} L. Franci, S. Landi, A. Verdini, L. Matteini and P. Hellinger, Astrophys. J. \textbf{853}, 26 (2018).

\bibitem{Carbone90} V. Carbone, P. Veltri and A. Mangeney, Phys. Fluids A \textbf{2}, 1487 (1990).

\bibitem{Pucci2014} F. Pucci and M. Velli, Astrophys. J. Lett. \textbf{780}, L19 (2014).

\bibitem{Comisso2016} L. Comisso, M. Lingam, Y.-M. Huang and A. Bhattacharjee, Phys. Plasmas \textbf{23}, 100702 (2016).

\bibitem{Tenerani2016} Tenerani, A., Velli, M., Pucci, F., Landi, S. and Rappazzo, A.F. Journal of Plasma Physics, \textbf{82}(5) (2016).

\bibitem{Huang2017} Y.-M. Huang, L. Comisso and A. Bhattacharjee, Astrophys. J. \textbf{849}, 75 (2017). 

\bibitem{Comisso2017} L. Comisso, M. Lingam, Y.-M. Huang and A. Bhattacharjee, Astrophys. J. \textbf{850}, 142 (2017). 

\bibitem{animation} See Supplemental Material \url{https://www.dropbox.com/s/2crexr8nu49ii0m/no_small_box.gif?dl=0} for the evolution of plasmoids in multiple intense current sheets.

\bibitem{Note1} \added{It is noteworthy that various theoretical predictions for accurate values of $k_*$ were all derived for forced turbulence, whereas in our simulations we consider decaying turbulence. In addition, the existent theoretical predictions do not take intermittency into account.}

\bibitem{Mallet2017} A. Mallet, A.A. Schekochihin and B.D.G. Chandran, MNRAS \textbf{468}, 4862 (2017).

\bibitem{LouBold2017} N.F. Loureiro and S. Boldyrev, Phys. Rev. Lett. \textbf{118}, 245101 (2017).

\bibitem{BoldLou2017} S. Boldyrev and N.F. Loureiro, Astrophys. J. \textbf{844}, 125 (2017).

\bibitem{Comisso2018} L. Comisso, Y.-M. Huang, M. Lingam, E. Hirvijoki and A. Bhattacharjee, Astrophys. J. \textbf{854}, 103 (2018). 

\bibitem{Toth2012} G. Toth et al., J. Comput. Phys. \textbf{231}, 870 (2012).

\bibitem{Mason2006} J. Mason, F. Cattaneo, and S. Boldyrev, Phys. Rev. Lett. \textbf{97}, 255002 (2006).

\bibitem{Greco2008} A. Greco, P. Chuychai, W. H. Matthaeus, S. Servidio, and P. Dmitruk, Geophys. Res. Lett.  \textbf{35}, L19111 (2008).

\bibitem{Iroshnikov63} P.S. Iroshnikov, Astron. Zh. \textbf{40}, 742 (1963).

\bibitem{Kraichnan65} R.H. Kraichnan, Physics of Fluids \textbf{8}, 1385 (1965).

\bibitem{Note2} \added{We also examined the kinetic energy spectra, where the incompressible and compressible components exhibit similar spectral shapes (not shown here), and manifest a clear ``break'' in the spectral index near $k=k_*$.}

\bibitem{Boldyrev2006} S. Boldyrev, Phys. Rev. Lett. \textbf{96}, 115002 (2006).


\end{thebibliography}
\end{document}